\pdfoutput=1

\documentclass{appolb}

\usepackage{amssymb}
\usepackage{hyperref}
\usepackage{graphicx}
\usepackage{amsmath}
\usepackage[numbers,sectionbib,sort&compress]{natbib}

\begin{document}

\title{Hollowness in $pp$ scattering%
\thanks{Talk presented by WB at XXIII Cracow EPIPHANY Conference, 9-12 January 2017.}
\thanks{Supported by Polish National Science Center grant 2015/19/B/ST2/00937, by Spanish Mineco Grant FIS2014-59386-P, 
and by Junta de Andaluc\'{\i}a grant FQM225-05.}}

\author{Wojciech Broniowski$^{1,2}$\thanks{Wojciech.Broniowski@ifj.edu.pl} and Enrique Ruiz Arriola$^3$\thanks{earriola@ugr.es}
\address{${}^{1}$The H. Niewodnicza\'nski Institute of Nuclear Physics, \\ Polish Academy of Sciences, PL-31342~Cracow, Poland}
\address{${}^{2}$Institute of Physics, Jan Kochanowski University, PL-25406~Kielce, Poland}
\address{${}^{3}$Departamento de F\'isica At\'omica, Molecular y Nuclear and \\ Instituto Carlos I de Fisica Te\'orica y Computacional,  Universidad de Granada, E-18071 Granada, Spain}}

\maketitle

\begin{abstract}
It is argued that the hollowness effect (depletion in the absorptive part of the scattering cross section at small values of the impact parameter) in the proton-proton 
scattering at the the LHC energies finds its origin in the quantum nature of the process, resulting in large values of the real part of the eikonal phase.
The effect cannot be reconciled with an incoherent superposition of the absorption from the proton constituents, thus suggests the change of this
basic paradigm of high-energy scattering.
\end{abstract}


\bigskip \bigskip

In this talk we discuss the significance of the recent $pp$ scattering
results from the Large Hadron Collider for our understanding of the
underlying physical processes in highest-energy collisions. In
particular, we argue that the {\em hollowness} in the inelastic cross
section treated as a function of the impact parameter $b$, i.e., its
depletion at low $b$, must necessarily originate from quantum
coherence, precluding a probabilistic folding interpretation.  More
details of our analysis can be found
in~\cite{Arriola:2016bxa,RuizArriola:2016ihz}, where we also analyze
the effect in 3-dimensions via the optical potential interpretation.

The TOTEM~\cite{Antchev:2013gaa} and ATLAS (ALFA)~\cite{Aad:2014dca}
Collaborations have measured the differential elastic cross section
for the $pp$ collisions at $\sqrt{s}=7$~TeV, later repeated for
$\sqrt{s}=8$~TeV~\cite{Antchev:2013paa,Aaboud:2016ijx}.  When the data
are used to obtain the inelastic cross section in the impact-parameter
representation, a striking feature appears: there is more inelasticity
when the two protons are separated by about half a fermi in the
traverse direction than for the head-on collisions. We term this
phenomenon {\em hollowness}. This unusual feature has been brought up
and interpreted by other
authors~\cite{Alkin:2014rfa,Dremin:2014eva,Dremin:2014spa,Dremin:2016ugi,Dremin:2017ylm,Anisovich:2014wha,Troshin:2016frs,Troshin:2017zmg,Troshin:2017ucy}.
A model realization of the effect was implemented via hot-spots
in~\cite{Albacete:2016pmp}.

\begin{figure}
\begin{center}
\includegraphics[width=0.51\textwidth]{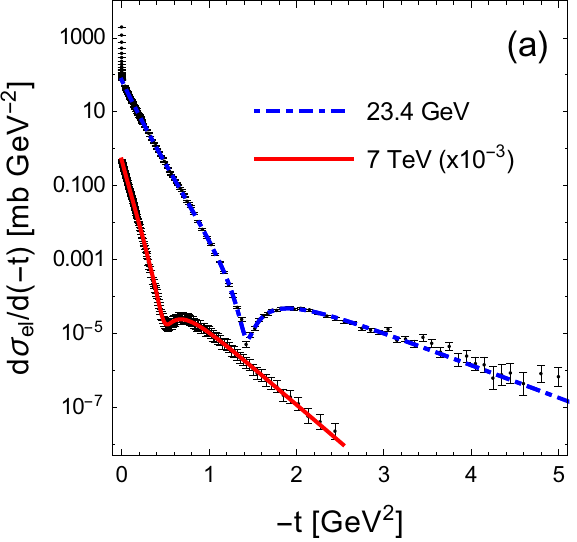} \hfill  \includegraphics[width=0.48\textwidth]{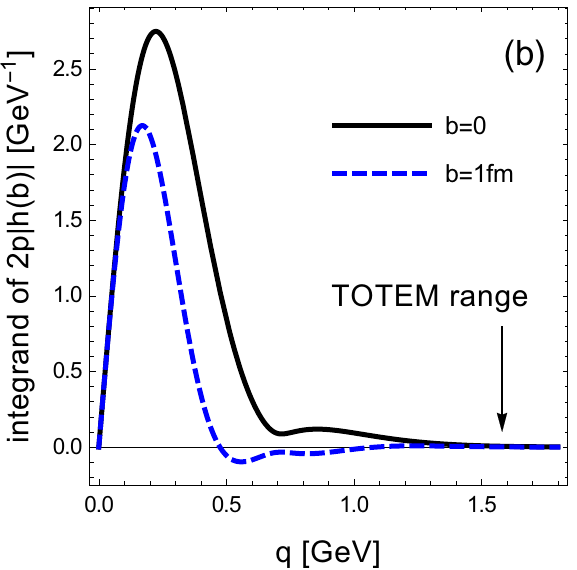} 
\end{center}
\vspace{-3mm}
\caption{(a) The data for the ISR energy of $\sqrt{s}=23.4$~GeV~\cite{Amaldi:1979kd} and the LHC energy 
of  $\sqrt{s}=7$~TeV~\cite{Antchev:2013gaa} with overlaid fits according to Eq.~\ref{eq:mBP2}. 
(b)~Plot of the integrand of Eq.~(\ref{eq:invf}), showing that that the range of the experimental data in $q$ is sufficient to carry out the 
Fourier-Bessel transform for the values of $b$ of interest. \label{fig:data}}
\end{figure}

We use the parametrization of the $pp$ scattering data~\cite{Fagundes:2013aja} based on the
Barger-Phillips model (modified BP2)~\cite{Phillips:1974vt}, with the form
\begin{eqnarray}
{\cal A} (s,t) &\equiv& \frac{f(s,t)}{p} = 
\sum_{n} c_n(s) F_n(t) s^{\alpha_n (t)} =\frac{i \sqrt{A} e^{\frac{B t}{2}}}{\left(1-\frac{t}{t_0}\right)^4}+i \sqrt{C} e^{\frac{D t}{2}+i \phi }, 
\nonumber \\ \label{eq:mBP2}
\end{eqnarray}
where $f(s,t)$ is the quantum mechanical scattering amplitude. The
modified BP2 model deals with the $t$ dependence, and the
$s$-dependent parameters are fitted separately to the differential
elastic $pp$ cross sections at $\sqrt{s}= 23.4$, $30.5$, $44.6$,
$52.8$, $62.0$, and $7000~{\rm GeV}$. A typical quality of
the fit, from the ISR~\cite{Amaldi:1979kd} at $\sqrt{s}=23.4$~GeV to
the LHC at $\sqrt{s}=7$~TeV, can be appreciated from
Fig.~\ref{fig:data}(a). These fits are not sensitive to
the {\it phase} of the scattering amplitude.

The $\rho(s)$ parameter is defined as the ratio of the real to imaginary parts of the 
amplitude at $t=0$:
\begin{eqnarray}
\rho(s) = \frac{{\rm Re}{\cal A} (s,0) }{{\rm Im}{\cal A} (s,0)}
\end{eqnarray}
This parameter has been recently determined for the LHC energy of
$\sqrt{s}=8$~TeV in~\cite{Antchev:2016vpy}.  To agree with this
experimental constraint we replace the parametrization of the
scattering amplitude of Eq.~(\ref{eq:mBP2}) with
\begin{eqnarray}
{\cal A} (s,t) \to \frac{i+\rho(s)}{\sqrt{1+\rho(s)^2}}|{\cal A} (s,t)|. \label{eq:rhoind}
\end{eqnarray}
This procedure assumes a $t$-independent ratio of the real to
imaginary parts of the scattering amplitude for all $t$-values, which
is the simplest choice. More general prescriptions have been analyzed
in detail in Ref.~\cite{Antchev:2016vpy}.  Our results presented below
are similar if we take, e.g., the Bailly et al.~\cite{Bailly:1987ki}
parametrization $\rho(s,t)=\rho_0(s)/(1-t/t_0(s))$, where $t_0(s)$ is
the position of the diffractive minimum.  However, admittedly, there
is some dependence on the choice of the model of
$\rho(s,t)$. Moreover, the problem is linked to the separation of the
Coulomb and strong amplitudes. The issue is crucial for the proper
extraction of the physical results and the ambiguity has a long
history since the early diagrammatic work of West and
Yennie~\cite{West:1968du}, which is consistent with the eikonal
approximation~\cite{Cahn:1982nr,Block:1984ru} but becomes sensitive to
internal structure from electromagnetic information such as form factors
(see, e.g.,~\cite{Prochazka:2016wno} and references therein).

\begin{table}[b]
\caption{Basic scattering observables for several collision energies 
obtained from Eq.~(\ref{eq:rhoind}), compared to experimental vales (lower rows). 
$B$ is the slope parameter of the differential elastic cross section. \medskip}
\begin{tabular}{cccccc}
 \hline
$\sqrt{s}$~[GeV] & $\sigma_{\rm el}$~[mb] & $\sigma_{\rm in}$~[mb] & $\sigma_{\rm T}$~[mb] & $B~[{\rm GeV}^{-2}]$ & $\rho$ \\ 
  \hline
 23.4 & 6.6 & 31.2 & 37.7 & 11.6 & 0.00 \\ 
\cite{Amaldi:1979kd} & 6.7(1) & 32.2(1) & 38.9(2) & 11.8(3) & 0.02(5) \\
  \hline
 200 & 10.0 & 40.9 & 50.9 & 14.4 & 0.13 \\ 
\cite{Aielli:2009ca,Bueltmann:2003gq} &  &  & 54(4) & 16.3(25) &  \\
  \hline
 7000 &  25.3 & 73.5 & 98.8 & 20.5 & 0.140 \\ 
\cite{Antchev:2013gaa} & 25.4(11) & 73.2(13) & 98.6(22)  & 19.9(3) & 0.145(100) \\
  \hline  
\end{tabular}
\label{tab:Fag-Mod}
\end{table}

Our prescription~(\ref{eq:rhoind}) maintains by construction the
quality of the fits shown in Fig.~\ref{fig:data}, but also the
experimental values for $\rho(s)$ are reproduced, which would not be
the case if Eq.~(\ref{eq:mBP2}) were used.  Basic physical quantities
stemming from our method are listed in Table~\ref{tab:Fag-Mod}, with
good agreement with the data supporting the used parametrization.

We now recall the relevant formulas from scattering theory: The $pp$
elastic differential cross section is given by
\begin{eqnarray}
\frac{d\sigma_{\rm el}}{dt}= \frac{\pi}{p^2} \frac{d \sigma_{\rm el}}{d \Omega} = 
\frac{\pi}{p^2} |f(s,t) |^2 = \pi | {\cal A} (s,t) |^2\,  , 
\end{eqnarray} 
with $p= \sqrt{s/4-M^2}$ the CM momentum and the partial wave
expansion of the scattering amplitude (we neglect spin effects) equal
to
\begin{eqnarray}
f(s,t) =\sum_{l=0}^ \infty (2l+1) f_l(p) P_l(\cos \theta).  \label{eq:PWA}
\end{eqnarray} 
The total cross section is given by the optical theorem, $\sigma_T =
4 \pi {\rm Im} f(s,0)/p$, and Coulomb effects are negligible at $|t| >
8 \pi \alpha /\sigma_T$, where $\alpha \simeq 1/137$ is the QED fine
structure constant and $\sigma_T$ is the total strong scattering cross
section. For $p a \gg 1$, with $a$ denoting the interaction range, one
can use the eikonal approximation with $bp = l +1/2 + {\cal
O}(s^{-1})$, where $b$ is the impact parameter. The $b$ representation
the scattering amplitude can be straightforwardly obtained from a
Fourier-Bessel transform of $f(s,t)$, known from the data
parametrization. Explicitly,
\begin{eqnarray}
2ph(b,s)={i} \left [ 1-e^{i \chi(b)}  \right ] =2p f_l(p) + {\cal O}(s^{-1}) = 2 \int_0^\infty q dq J_0(bq) f(s,-q^2).
\nonumber \\ \label{eq:invf}
\end{eqnarray} 
In Fig.~\ref{fig:data}(b) we demonstrate  that the range of the TOTEM data in $q$ is sufficient to carry out this transform to  
a satisfactory accuracy needed in our analysis. 

\begin{figure}
\begin{center}
\includegraphics[width=0.49\textwidth]{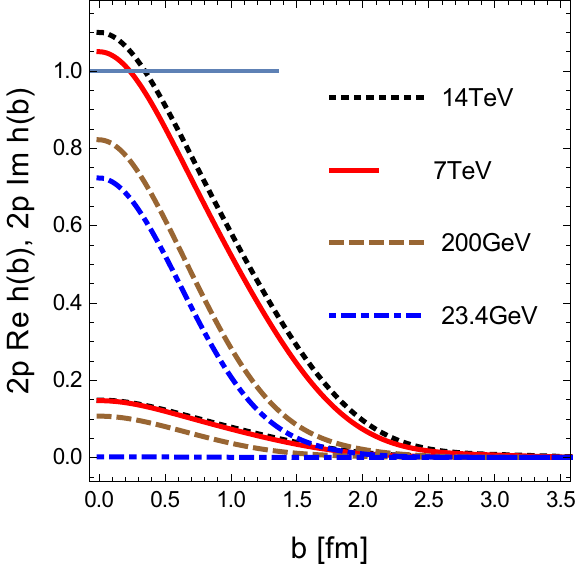} 
\end{center}
\vspace{-3mm}
\caption{Real (lower curves) and imaginary (upper curves) parts of the  eikonal scattering amplitude $2ph(b)$ for several collision energies. 
We note that for the LHC energies, 
at the origin $2p {\rm Im} h(0)>1$. \label{fig:amp}}
\end{figure}  

\begin{figure}
\begin{center}
\includegraphics[width=0.49\textwidth]{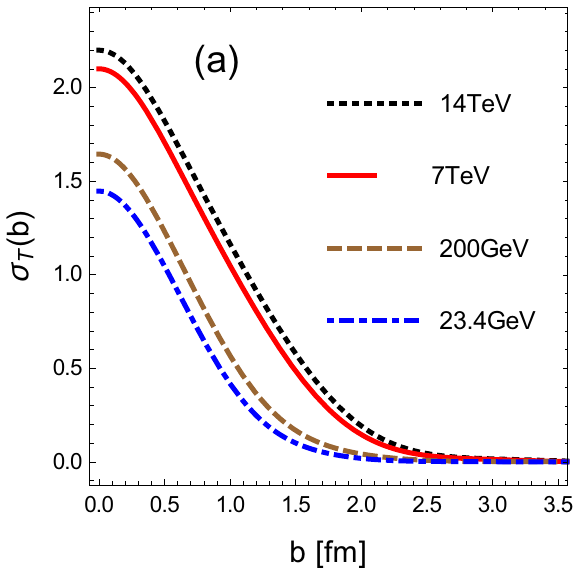} \hfill \includegraphics[width=0.49\textwidth]{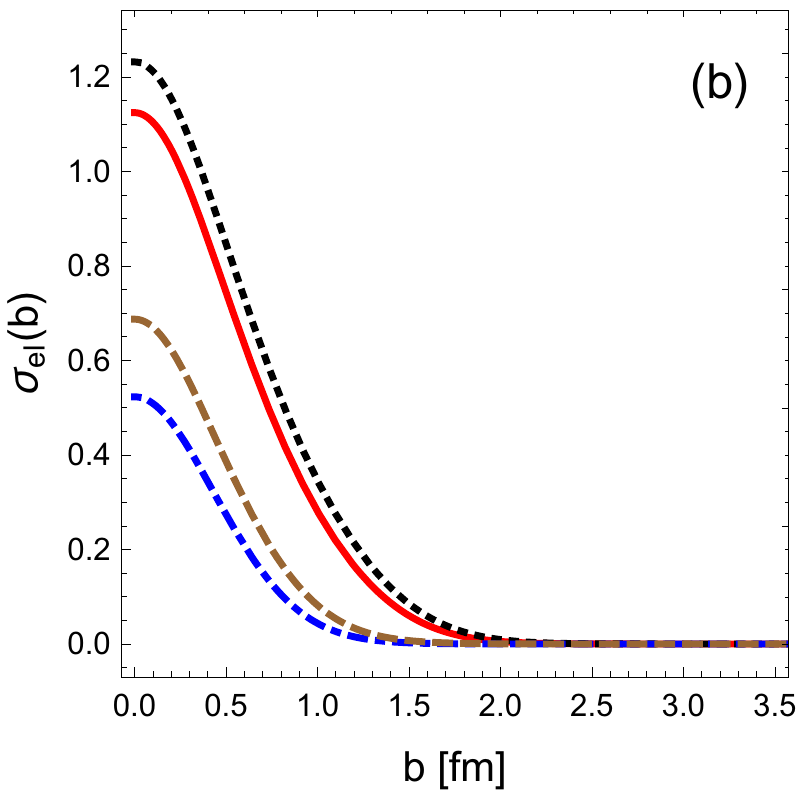} \\
\includegraphics[width=0.49\textwidth]{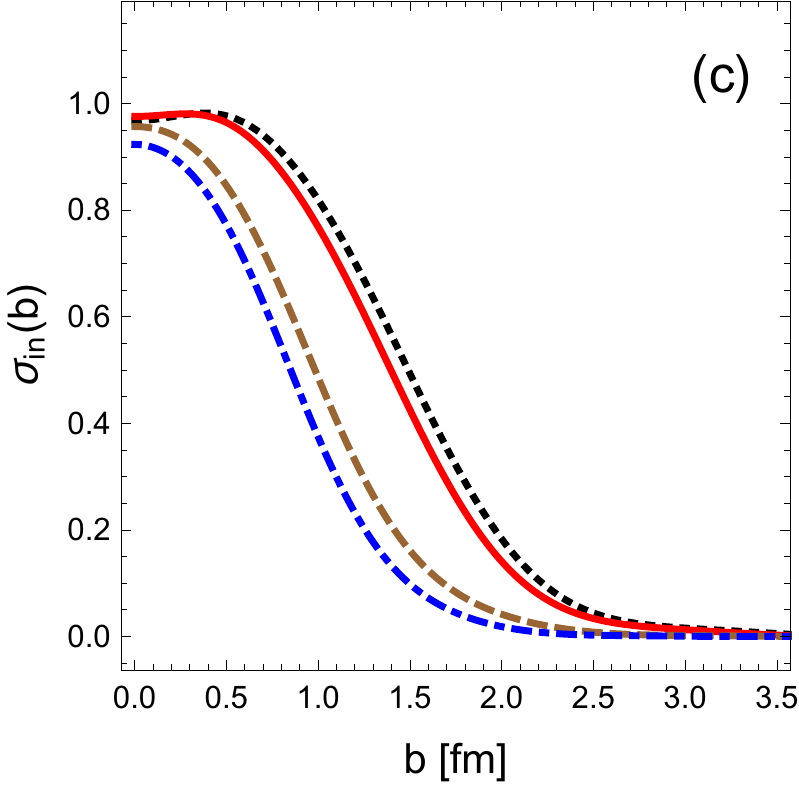} \hfill \includegraphics[width=0.501\textwidth]{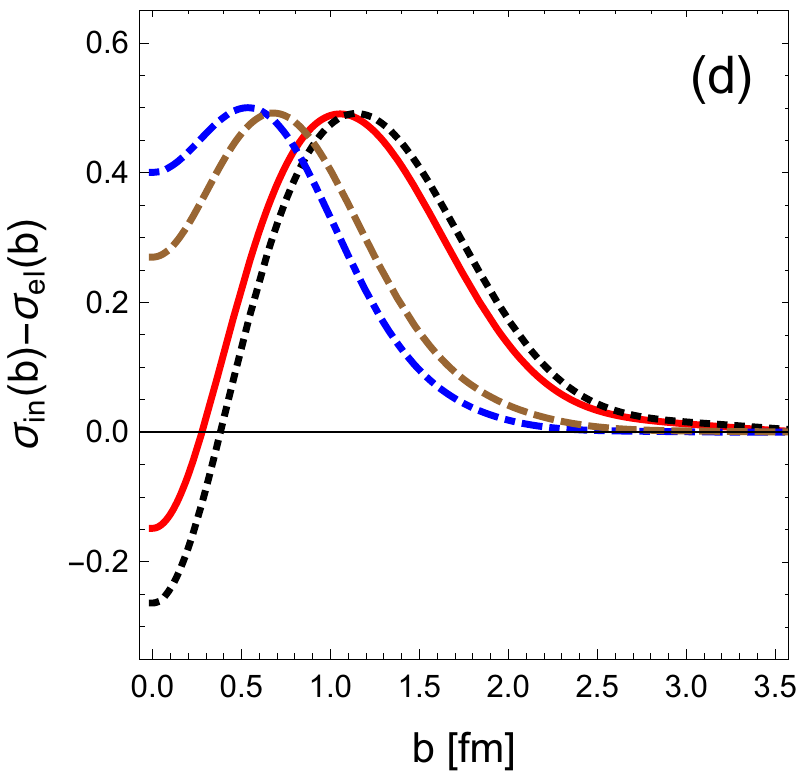} \\
\end{center}
\vspace{-3mm}
\caption{The total (a), elastic (b), and inelastic (c) cross section, as well as the edge function, plotted as functions of
the impact parameter at various collision energies. \label{fig:sigmas}}
\end{figure}

The standard formulas for the total, elastic, and inelastic cross sections (in our analysis we treat all the components to the  inelastic 
scattering jointly, not discriminating, e.g.,  the diffractive components) in the $b$ representation can be parameterized with the eikonal phase $\chi(b)$
and have the form~\cite{Blankenbecler:1962ez}
\begin{eqnarray} 
\sigma_T  &=& \frac{4 \pi}p {\rm Im} f(s,0) = 4 p \int d^2 b {\rm Im} h(\vec b,s) = 2 \int d^2 b \left [ 1- {\rm Re} \, e^{i \chi(b)} \right ] \, \label{eq:st} ,\\
\sigma_{\rm el} &=& \int d\Omega |f(s,t)|^2 = 4 p^2 \int d^2 b |h(\vec b,s)|^2 =  \int d^2 b | 1- e^{i \chi(b)} |^2 \, \label{eq:sel}, \\
\sigma_{\rm in} &\equiv& \sigma_T - \sigma_{\rm el} = \int d^2 b \sigma_{\rm in} (b) =  \int d^2 b \left [ 1- e^{- 2 {\rm Im} \chi (b)} \right ] ,  \label{eq:sin}
\end{eqnarray} 
with the integrands $\sigma_{\rm in} (b)$, $\sigma_{\rm el} (b)$ and
$\sigma_{\rm T} (b)$ being dimensionless quantities that can be
interpreted as the corresponding $b$-dependent relative number of
collisions.  For instance, accordingly to Eq.~(\ref{eq:sin}), the
inelasticity profile is defined as
\begin{eqnarray}
\sigma_{\rm in} (b)  = 4p {\rm Im} h(b,s) -  4p^2|h(b,s)|^2.   \label{eq:prof}
\end{eqnarray} 
While unitarity implies $\sigma_{\rm in}(b)> 0$, one also has
$\sigma_{\rm in}(b) \le 2 k(b,s) - k(b,s)^2$, with $k(b,s) \equiv 2 p
{\rm Im h(b,s)}$, and hence one also has the upper bound \mbox{$\sigma_{\rm
in}(b) \le 1$}.

\begin{figure}
\begin{center}
\includegraphics[width=0.49\textwidth]{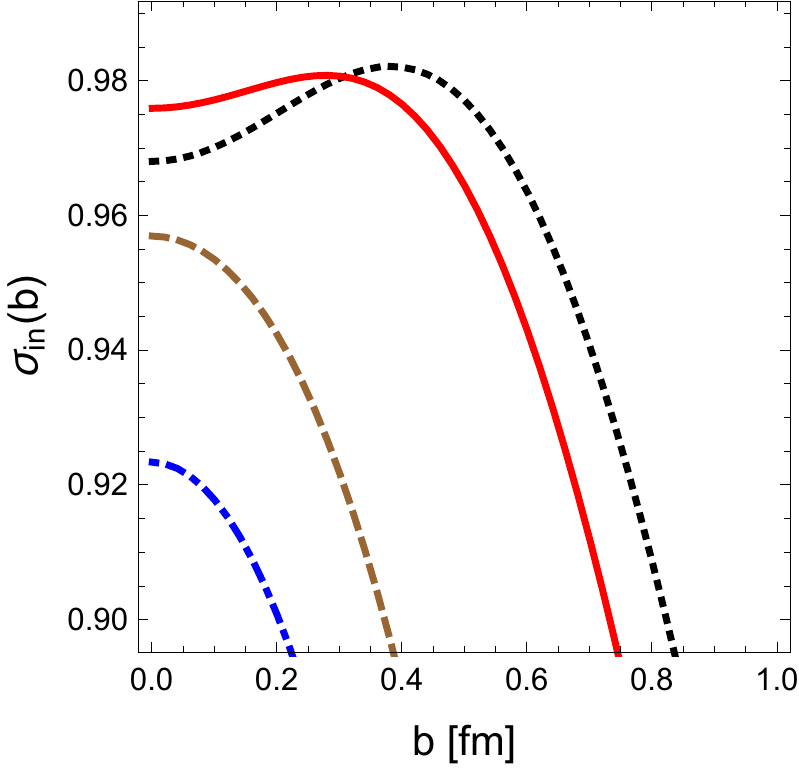} 
\end{center}
\vspace{-3mm}
\caption{A close-up of Fig.~\ref{fig:sigmas}(c). \label{fig:closeup}}
\end{figure}

Now we come to our results. In Fig.~\ref{fig:amp} we present the real and imaginary parts of the eikonal amplitude $2ph(b)$ for several collision 
energies. The real parts are smaller from the corresponding imaginary parts, as their ratio is given by the (constant) $\rho$ parameter. 
The important observation here is that the imaginary parts go above 1 near the origin for the LHC collision energies. We will come back to this issue shortly.
  
\begin{figure}
\begin{center}
\includegraphics[width=0.49\textwidth]{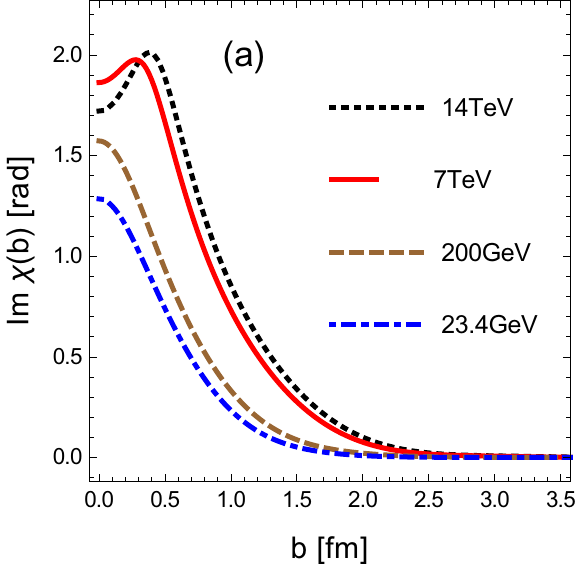} \hfill \includegraphics[width=0.49\textwidth]{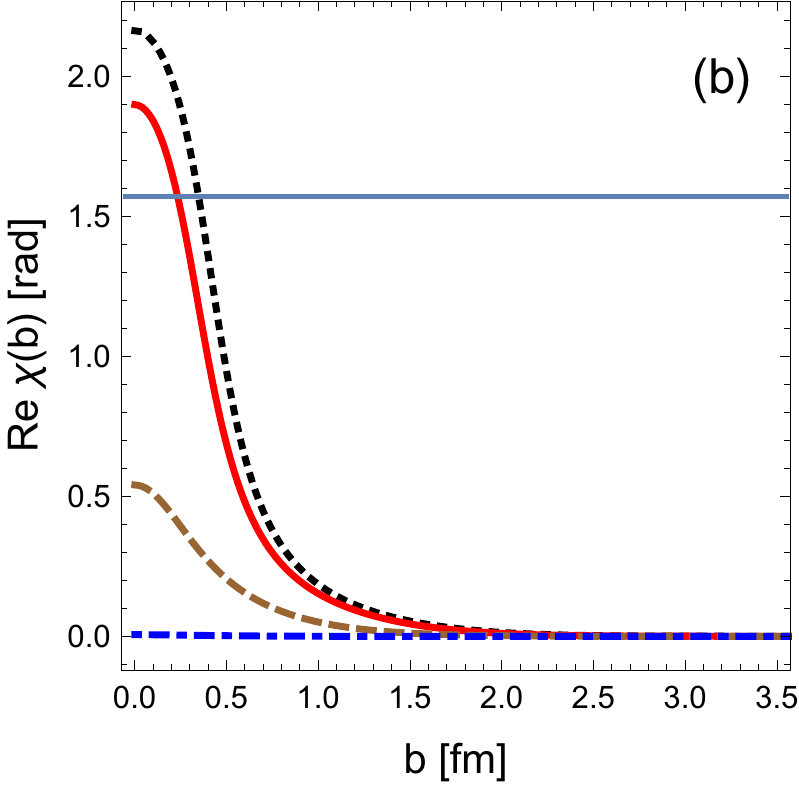}
\end{center}
\vspace{-3mm}
\caption{Imaginary (a) and real (b) part of the eikonal scattering phase, plotted as functions of the impact parameter for several collision 
energies. We note that at the LHC energies ${\rm Re}\chi(b=0)$ goes above $\pi/2$. \label{fig:chi}}
\end{figure}  
 
In Fig.~\ref{fig:sigmas} we collect the results for the impact-parameter representations of the total, elastic, and inelastic cross sections, as well as 
for the {\em edge} function~\cite{Block:2014lna,Block:2015sea}, defined as $\sigma_{\rm in}(b)-\sigma_{\rm el}(b)$. 
The most important feature, visible from Fig.~\ref{fig:sigmas} and more accurately form the close-up of Fig.~\ref{fig:closeup}, is the {\em hollowness}: 
the inelastic cross section develops a minimum at $b=0$ at the LHC collision energies. 

To better understand these results, one should resort to the 
formulas expressed with the eikonal phase, plotted in Fig.~\ref{fig:chi}. We have
\begin{eqnarray}
2p{\rm Im}\, h(b) &=& 1-e^{-{\rm Im}\chi(b)}\cos {\rm Re}\chi(b), \label{eq:eik2} \\
2p{\rm Re}\, h(b) &=& e^{-{\rm Im}\chi(b)}\sin {\rm Re}\chi(b), \nonumber \\
\sigma_T(b)&=& 2 - 2 e^{-{\rm Im}\chi(b)} \cos {\rm Re}\chi(b), \nonumber \\ 
\sigma_{\rm el}(b)&=& 1 + e^{-2 {\rm Im}\chi(b)} - 2 e^{-{\rm Im}\chi(b} \cos {\rm Re}\chi(b), \nonumber \\ 
\sigma_{\rm in}(b)&=& 1 - e^{-2 {\rm Im}\chi(b)}, \nonumber \\ 
\sigma_{\rm el}(b)-\sigma_{\rm in}(b)&=&  2 e^{-{\rm Im}\chi(b)} \left[ \cos{\rm Re} \chi(b)-e^{-{\rm Im}\chi(b)} \right]. \nonumber
\end{eqnarray}
We note several facts following from the above relations:
\begin{enumerate}
 \item Going of $2p{\rm Im}\, h(b)$ above 1 and $\sigma_T(b)$ above 2
 are caused by ${\rm Re} \chi(b)>\pi/2$, where $\cos {\rm Re}\chi(b)<0$
 (cf.~Figs.~\ref{fig:amp}, \ref{fig:sigmas}(a), and \ref{fig:chi}(b)).  \item In addition,
 if $\chi(b)>\pi/2$, the edge function is negative and $\sigma_{\rm
 el}(b)>1$.  \item The departure of $2p{\rm Im}\, h(b)$ from 1 is of
 similar order as $2p{\rm Re}\, h(b)$, with both suppressed with
 $e^{-{\rm Im}\chi(b)}$.
\end{enumerate}
We see that this is the {\em real} part of the eikonal phase which
controls the behavior related to hollowness.

\begin{figure}
\begin{center}
\includegraphics[width=0.49\textwidth]{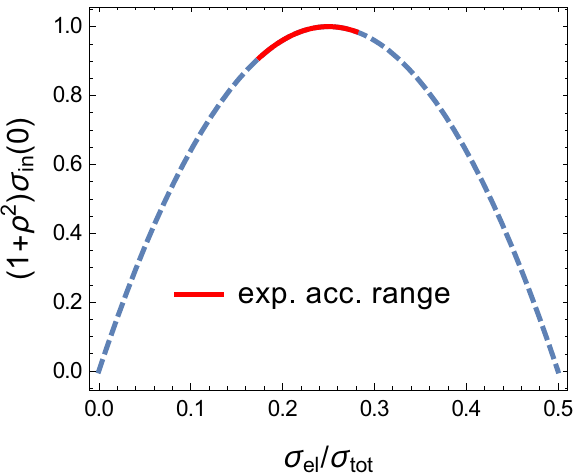} 
\end{center}
\vspace{-3mm}
\caption{Illustration of Eq.~(\ref{eq:ga}). The solid line corresponds to the experimental values of the 
ratio of the elastic to total $pp$ cross section. Values of
$\sigma_{\rm el}/\sigma_T>1/4$ correspond to hollowness in the
Gaussian model. \label{fig:parab}}
\end{figure}  

One may give a simple criterion for $\sigma_{\rm in}(b)$ to develop
a minimum at $b=0$. 
From Eqs.~(\ref{eq:prof}) and (\ref{eq:rhoind}) we get
\begin{eqnarray}
\frac{d\sigma_{\rm in}(b)}{db^2}= 2p \frac{d{\rm Im}\, h(b)}{db^2} \left [ 1-(1+\rho^2) 2p {\rm Im}\, h(b) \right ],
\end{eqnarray}
which is negative at the origin if
\begin{eqnarray}
2 p {\rm Im} h (0)>\frac{1}{1+\rho^2}\sim 1. \label{eq:crit}
\end{eqnarray}
Since $\rho=0.14$ at the LHC, the departure of $1/(1+\rho^2)$ from 1
is at a level of $2\%$.

We also find from Eq.~(\ref{eq:eik2}) that 
\begin{eqnarray}
 \frac{d\sigma_{\rm in}(b)}{db^2} = 2  e^{-2 {\rm Im}\chi(b)} \frac{d {\rm Im}\chi(b)}{db^2},
\end{eqnarray}
thus the appearance of the dip at the origin in $\sigma_{\rm in}(b)$  is associated with the dip in ${\rm Im}\chi(b)$. 
This is manifest between Fig.~\ref{fig:closeup} and Fig.~\ref{fig:chi}(a). 

Dremin~\cite{Dremin:2014eva,Dremin:2014spa,Dremin:2016ugi} proposed a simple Gaussian model of the amplitude which one may adapt to 
the presence of the real part of the amplitude (which is crucial for maintaining unitarity with the hollowness effect). One can 
parametrize the amplitude at low values of $b$ (which is the numerically relevant region) as
\begin{eqnarray}
&& {{\rm Im}(2p \, h(p)) = A e^{-\frac{2b^2}{2B}} }, \;\; 
A=\frac{4 \sigma_{\rm el}}{\left(1+\rho^2\right) \sigma_{\rm tot}}, \;\; B=\frac{\left(1+\rho^2\right) \sigma_{\rm tot}^2}{16 \pi  \sigma_{\rm el}}.
\end{eqnarray}
The curvature of the inelasticity profile at the origin is
\begin{eqnarray}
\frac{1}{2} \left . \frac{d^2 n_{\rm in}(b)}{db^2} \right |_{b=0} = 
 \frac{64 \pi \sigma_{\rm el}^2 (4\sigma_{\rm el}- \sigma_{\rm
 tot})}{\left(\rho^2+1\right)^2 \sigma_{\rm tot}^4},
\end{eqnarray}
We note it changes sign when {$\sigma_{\rm el}=\frac{1}{4}\sigma_{\rm tot}$}, 
with the value at the origin 
\begin{eqnarray}
\sigma_{\rm in}(0)= \frac{8\sigma_{\rm el}}{(1+\rho^2) \sigma_{\rm tot}} \left ( 1 - 2 \frac{\sigma_{\rm el}}{\sigma_{\rm tot}} \right). \label{eq:ga}
\end{eqnarray}
As predicted by Dremin, the hollowness effect emerges when  $\sigma_{\rm el}>\frac{1}{4}\sigma_{\rm tot}$, which 
is the case of the LHC collision energies. We illustrate relation~(\ref{eq:ga}) in Fig.~\ref{fig:parab}. 

The final point, very important from the conceptual point of view and
for the understanding of the effect, is the impossibility of hollowness to emerge
from incoherent folding of inelasticities of collisions of the
protons' partonic constituents.  In many models incoherent
superposition is assumed, i.e., the inelasticity of the $pp$ process
is obtained from the folding formula shown below. These ideas have been implemented in
microscopic models based on intuitive geometric
interpretation~\cite{Chou:1968bc,Chou:1968bg,Cheng:1987ga,Bourrely:1978da,Block:2006hy,Block:2015sea}. 
Folding involves
\begin{eqnarray}
\sigma_{\rm in}(b) &\propto& \int d^2 b_1 d^2 b_2 
\rho( \vec b_1 + \vec b/2) w(\vec b_1 - \vec b_2 ) \rho(\vec b_2-\vec b/2) \nonumber \\ &=& \int d^3 b_1 d^3 b_2 \rho(\vec b_1) w(\vec b_1 - \vec b_2 ) \rho(\vec b_2)  \nonumber \\ 
&-& \frac12  \int d^3 b_1 d^3 b_2 [\vec b \cdot \nabla \rho(\vec b_1)]
w(\vec b_1-\vec b_2) [\vec b \cdot \nabla \rho(\vec b_2)]+ \dots,
\end{eqnarray}
where $w(\vec b_1-\vec b_2)$ is a positive-definite kernel (folding
models usually take $w(\vec b_1-\vec b_2) \propto \delta (\vec b_1
-\vec b_2)$) and $\rho(\vec{b})$ describes the (possibly correlated)
transverse distribution of components in the proton.  By passing to
the Fourier space it is simple to show that {$\sigma_{\rm
in}(b)=\alpha^2 - \beta^2 b^2+\dots$}, with real constants $\alpha$ and
$\beta$, therefore $\sigma_{\rm in}(b)$ has necessarily a {local
maximum} at $b=0$, in contrast to the phenomenological hollowness
result at the LHC energies. An analogous argument holds for the
3D-hollowness unveiled in our
work~\cite{Arriola:2016bxa,RuizArriola:2016ihz}, which takes place already at
lower energies.

\bigskip

In conclusion, we stress that the hollowness effect in $pp$ scattering at the LHC energies has necessarily  a quantum
origin. As just shown, it cannot be obtained by an incoherent folding
of inelasticities of collisions of partonic constituents. Moreover, we have demonstrated that the real part of the
scattering amplitude plays a crucial role in generating hollowness:
the effect appears when the real part of the eikonal phase becomes
larger than $\pi/2$. Per se, there is nothing unusual in that fact. If
coherence occurs, the phases of amplitudes from the constituents may
add up (as is the case, e.g., in the Glauber
model~\cite{glauber1959high}) and at some point the value of $\pi/2$
may be crossed. A microscopic realization of this quantum mechanism remains, however,
a challenge.  Finally, we note that
in~\cite{Arriola:2016bxa,RuizArriola:2016ihz} we have presented a
three-dimensional interpretation of the effect, which offers an even
more pronounced hollowness feature.

\bibliography{NN-high-energy}
 
\end{document}